\documentclass{article}
\usepackage[utf8]{inputenc}
\usepackage{amsmath,amsfonts,graphicx}
\usepackage{hyperref}
\usepackage{caption}
\title{Hierarchical Robust PCA for Scalable Data Quality Monitoring in Multi-level Aggregation Pipelines}
\author{Preetam Kumar Ojha \\
Netflix Inc. \\
\texttt{pojha@netflix.com}}
\date{}

\begin{document}

\maketitle

\begin{abstract}
Data quality (DQ) remains a fundamental concern in big data pipelines, especially when aggregations occur at multiple hierarchical levels. Traditional DQ validation rules often fail to scale or generalize across dimensions such as user interactions, sessions, profiles, accounts, and regions. In this paper, we present a novel application of Hierarchical Robust Principal Component Analysis (HrPCA) as a scalable, unsupervised anomaly detection technique tailored to DQ monitoring in multi-level aggregation pipelines. We propose a modular framework that decomposes the data at each hierarchical level into low-rank representations and sparse residuals, allowing the detection of subtle inconsistencies, outliers, and misalignments in the aggregated data. We evaluated our approach using synthetic hierarchical datasets with controlled anomalies and demonstrated how HrPCA outperforms traditional rule-based methods in detecting data corruption and rollup inconsistencies.
\end{abstract}

\section{Introduction}
Ensuring high-quality data is a growing challenge in complex data ecosystems where data is collected from numerous logging sources. Most of the logging occurs at the user interaction level, such as button clicks, scrolls, content plays, and other in-app behaviors, which are then aggregated to session-level summaries and further rolled up to profile-, account-, and region-level datasets. Often, these interaction logs contain correlated or partially redundant information that is not immediately obvious. Downstream systems aggregate and transform this data to produce business-critical metrics, dashboards, and machine learning features. When anomalies occur in the aggregated metrics, the underlying cause is not always apparent — especially if it is due to a subtle inconsistency in upstream logging behavior. In such cases, data engineers can spend hours manually inspecting logs, tracing transformations, and reverse engineering the behavior of the pipeline to locate the root issue. Current data quality tools can validate schemas and row counts, but they do not identify structural deviations in data that span multiple layers of aggregation. This gap makes scalable root cause analysis a major bottleneck.

Robust Principal Component Analysis (RPCA) has shown promise in separating structured signals from sparse corruptions in high-dimensional data. We extend this concept to a hierarchical setting, introducing Hierarchical Robust PCA (HrPCA) for the first time as a data quality audit mechanism. By modeling low-rank signals at each level of aggregation and isolating unexpected deviations as sparse anomalies, HrPCA provides a mathematically grounded, interpretable, and scalable solution for detecting DQ issues in pipelines.

\section{Related Work}
Prior work on data quality monitoring has focused on rule-based validation (e.g. Deequ, Great Expectations), metric comparison audits, and statistical anomaly detection. Although RPCA has been used in video surveillance and bioinformatics for outlier detection, its application in structured hierarchical data systems has been limited.

Our work extends RPCA to a modular form that aligns with real-world pipeline hierarchies, enabling decomposition and validation at each level without hand-crafted rules. This approach is inspired by the success of multiresolution decomposition in signal processing and structured low-rank recovery in computer vision.

\section{Methodology}
HrPCA decomposes the data from each hierarchical level into a low-rank structure and a sparse residual. Consider data organized at levels such as the session \(\rightarrow\) profile \(\rightarrow\) account. For each level \(i\), we represent the observed data matrix \(X_i\) as:

\[
X_i = L_i + S_i
\]

Where:
\begin{itemize}
    \item \(L_i\): Low-rank approximation capturing expected patterns at level \(i\)
    \item \(S_i\): Sparse matrix that captures corruptions, outliers, and anomalies
\end{itemize}

The total reconstruction is:
\[
\hat{X} = \sum_{i=1}^k L_i \quad \text{and} \quad S = X - \hat{X}
\]

We implement this using Truncated SVD for each level and retain only the top \(r_i\) components determined by explained variance or downstream constraints. Anomalies are flagged based on row-wise \(\ell_2\) norms of the sparse matrix exceeding a threshold.

\section{System Architecture}
The HrPCA-based DQ monitoring system is modular and scalable. Raw input tables (e.g., playback sessions, interaction logs) are aggregated into feature matrices per hierarchy level using Apache Spark. These matrices are versioned and passed into HrPCA audit modules.

\begin{figure}[h!]
\centering
\includegraphics[width=0.8\textwidth]{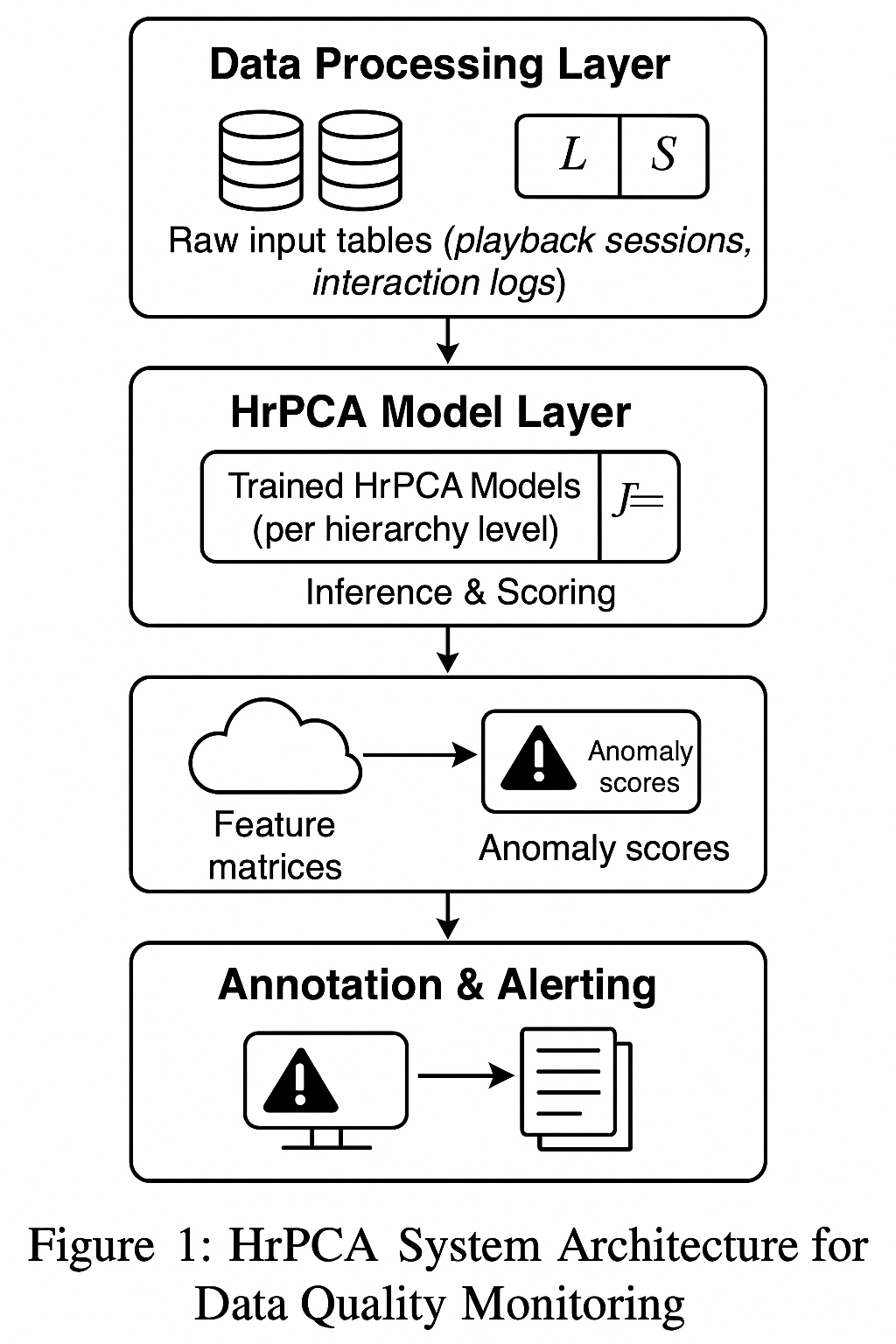}
\caption{HrPCA System Architecture for Data Quality Monitoring}
\label{fig:architecture}
\end{figure}

\begin{itemize}
    \item \textbf{Data Processing Layer}: Spark jobs extract feature vectors for each hierarchy level and partition by date.
    \item \textbf{HrPCA Model Layer}: The trained HrPCA models are stored for each hierarchy level and used for inference.
    \item \textbf{Inference \& Scoring}: Daily/hourly data are projected onto the learned low-rank bases; residuals are calculated.
    \item \textbf{Anomaly Flagging}: Sparse components that exceed dynamic thresholds are flagged.
    \item \textbf{Annotation \& Alerting}: Detected anomalies are correlated with product change logs or deployment timelines.
\end{itemize}

\section{Experiments}
We generate synthetic hierarchical datasets that mimic session, profile, and account aggregations. Clean low-rank data are constructed using randomized projections. Controlled anomalies are injected into random subsets of the data via additive noise. The HrPCA model is trained on clean data, and inference is performed on corrupted samples.

Visualizations include:
\begin{itemize}
    \item Heatmaps of residuals showing affected rows and features
    \item Line plots of anomaly scores with threshold demarcations
\end{itemize}

\begin{figure}[h!]
\centering
\includegraphics[width=0.8\textwidth]{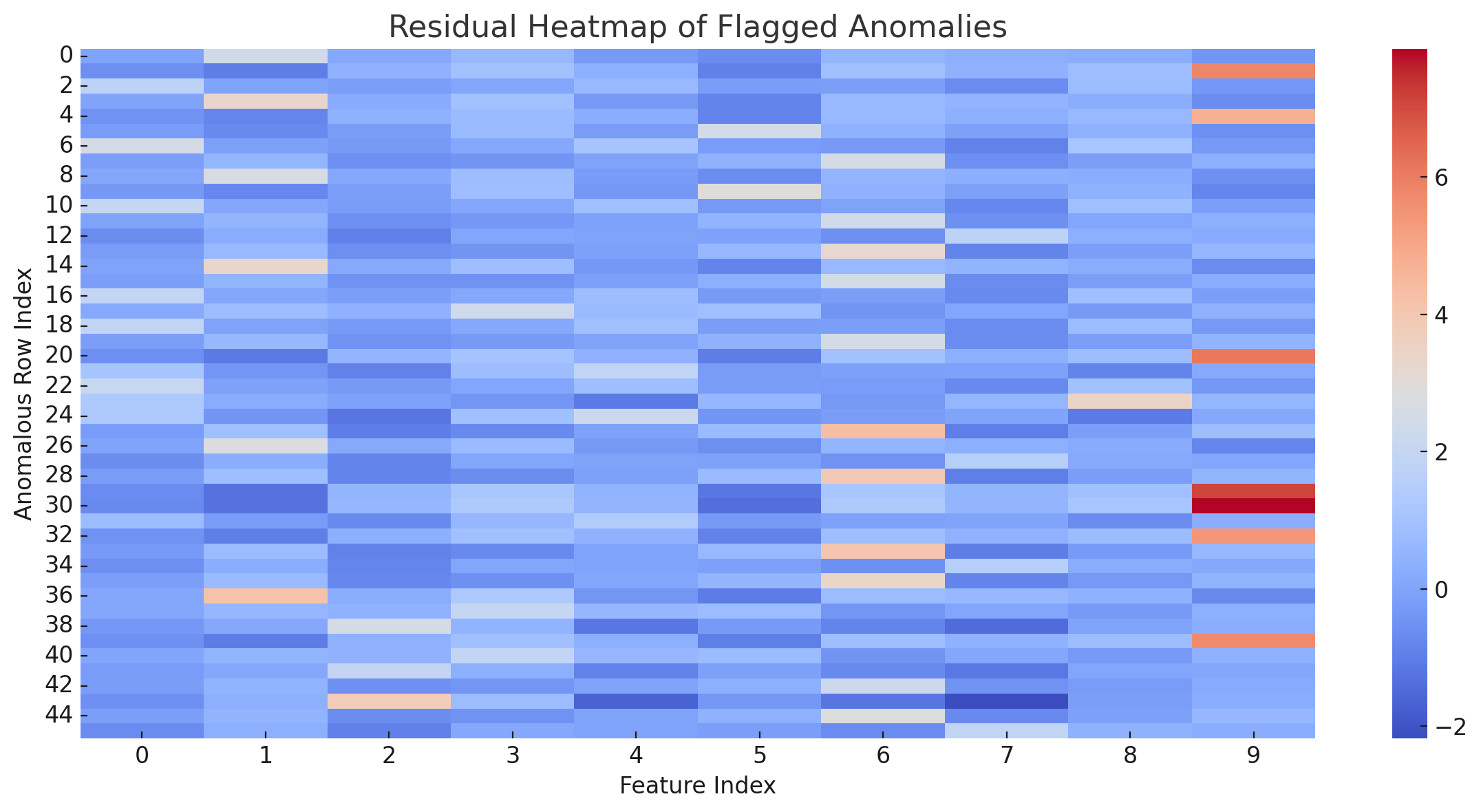}
\caption{Residual Heatmap}
\label{fig:Heatmaps of residuals showing affected rows and features}
\end{figure}

\begin{figure}[h!]
\centering
\includegraphics[width=0.8\textwidth]{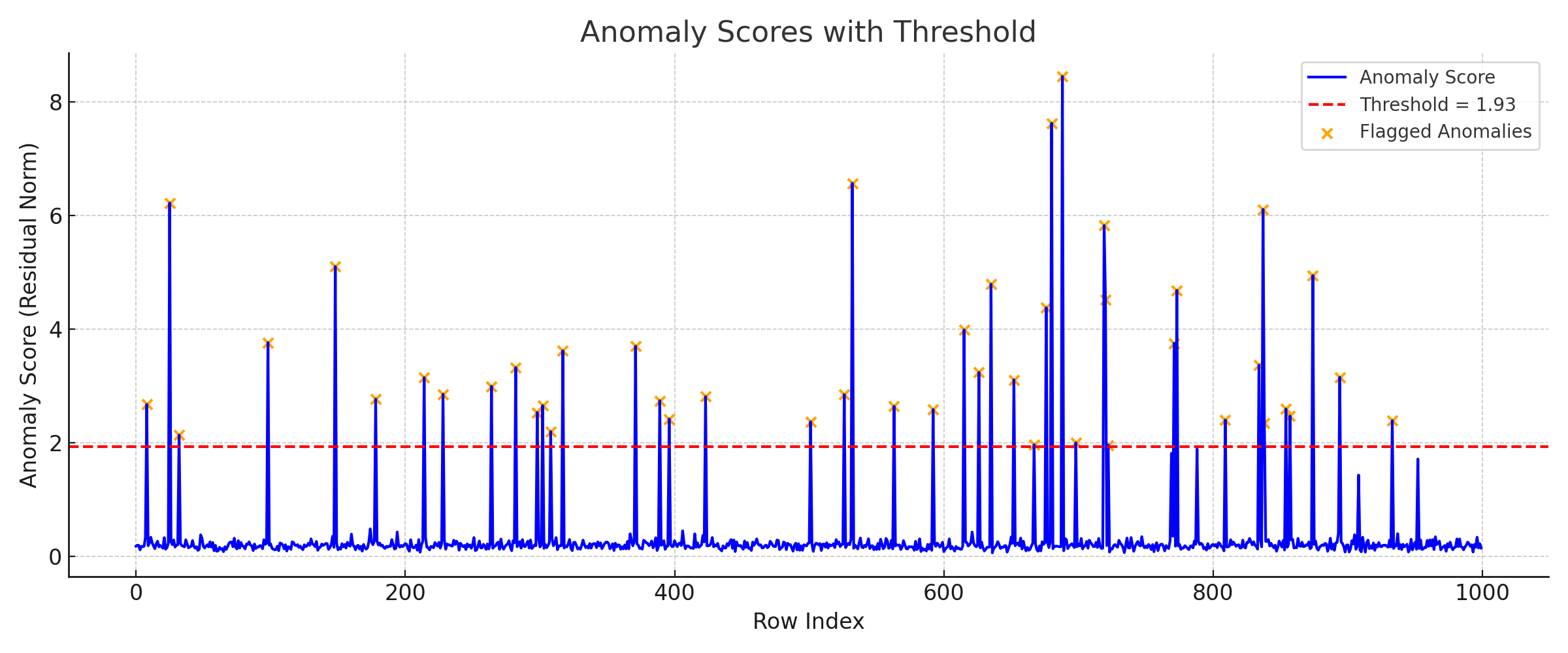}
\caption{Anomaly Score Plots}
\label{fig:Line plots of anomaly scores with threshold demarcations}
\end{figure}

Evaluation metrics include:
\begin{itemize}
    \item \textbf{Precision}: Ratio of true anomalies flagged to total flagged
    \item \textbf{Recall}: Ratio of true anomalies flagged to total injected
    \item \textbf{F1 Score}: Harmonic Mean of Precision and Recall
\end{itemize}

\subsection{Evaluation Metrics for Anomaly Detection}
To quantify the effectiveness of the HrPCA anomaly detection framework, we evaluate its performance using three standard classification metrics: precision, recall, and F1 score. These metrics are computed by comparing the predicted anomaly flags from HrPCA to the ground-truth anomaly labels in the synthetic dataset (injected during experiment generation).

Let:
\begin{itemize}
    \item $TP$ = True Positives: Number of correctly detected anomalies
    \item $FP$ = False Positives: Number of normal rows incorrectly flagged as anomalies
    \item $FN$ = False Negatives: Number of injected anomalies that went undetected
\end{itemize}

The metrics are then defined as:
\begin{equation}
\text{Precision} = \frac{TP}{TP + FP}
\end{equation}

\begin{equation}
\text{Recall} = \frac{TP}{TP + FN}
\end{equation}

\begin{equation}
\text{F1 Score} = 2 \cdot \frac{\text{Precision} \cdot \text{Recall}}{\text{Precision} + \text{Recall}}
\end{equation}

To validate HrPCA in a multilevel aggregation context, we simulated a synthetic hierarchical data set consisting of four levels: interaction level, session level (aggregated over 5 interactions), profile level (aggregated over 5 sessions) and account level (aggregated over 5 profiles). Each level had controlled anomalies injected into the lowest level and propagated upward.

Separate HrPCA models were trained and evaluated at each level using the same dimensionality ($n=1$) while sweeping over residual thresholds.

\begin{table}[h!]
\centering
\caption{HrPCA Performance Across Hierarchical Aggregation Levels}
\begin{tabular}{|l|c|c|c|c|}
\hline
\textbf{Level} & \textbf{Threshold} & \textbf{Precision} & \textbf{Recall} & \textbf{F1 Score} \\
\hline
Interaction & 5.24 & 1.00 & 1.00 & 1.00 \\
Session     & 2.02 & 1.00 & 1.00 & 1.00 \\
Profile     & 0.50 & 0.88 & 1.00 & 0.94 \\
Account     & 0.00 & 0.00 & 0.00 & 0.00 \\
\hline
\end{tabular}
\label{tab:hrpca_hierarchy}
\end{table}

As shown in Table~\ref{tab:hrpca_hierarchy}, HrPCA performs optimally at the lower levels of the hierarchy (interaction and session), where the anomaly signals are strongest. As data is aggregated to higher levels (profile and account), signal dilution reduces detection effectiveness. At the account level, anomaly signals are too diffuse to be detected using current configurations, highlighting the need for enhanced tracing or multilevel fusion strategies.

\section{Eigenvector Backtracking and Anomaly Annotation}

A key benefit of HrPCA is its interpretability. By analyzing the top eigenvectors from the decomposition of each level, we can trace which dimensions (features) contribute the most to anomalies. High projection weights along specific eigenvectors often correlate with structural changes or logging issues in the upstream pipeline.

Each eigenvector corresponds to a latent pattern or direction of variability in the data. Anomalous rows (e.g., session-level feature vectors) tend to have large projections onto specific eigenvectors when the corresponding pattern is broken. Formally, for a row vector \( x \) and an eigenvector \( u_j \), the projection score is given by:
\[
p_j = x^\top u_j
\]

If \( |p_j| \) is significantly higher than expected (based on historical norms or standard deviation thresholds), the direction \( u_j \) is considered to explain the anomaly.

To understand the contribution of each feature dimension to this projection, we can inspect the magnitude of each element in \( u_j \). The index \( i \) with the highest absolute value \( |u_{ji}| \) indicates the most influential feature that contributes to the anomaly.

This property enables several forms of root-cause reasoning:
\begin{itemize}
    \item \textbf{Feature attribution}: Use the decomposition \( x \approx \sum_j p_j u_j \) to analyze the projection components.
    \item \textbf{Temporal cross-referencing}: Track the time series of \( p_j(t) \) for each eigenvector to correlate spikes with deployments, A/B tests, or configuration changes.
    \item \textbf{Hierarchical tracing}: Since HrPCA models are trained per aggregation level (e.g., session, profile, account), backtracking can localize the issue's origin across the hierarchy.
    \item \textbf{Metric annotation}: Anomalies are tagged with the dominant eigenmode direction, enabling alerts to carry context such as "engagement volume deviation" or "sudden drop in unique sessions".
\end{itemize}

This backtracking not only flags the anomaly, but also provides a diagnostic trail back to the most likely feature group or structural reason behind the data issue. We believe that this makes HrPCA particularly effective in systems where data engineers and analysts need fast, actionable insight into breakdowns across complex data products.

By associating anomalies with known product changes, we close the loop between pipeline monitoring and the business context.

\section{Conclusion and Future Work}
HrPCA provides a scalable, interpretable, and reusable method for DQ auditing in big data pipelines. Its ability to operate across multiple levels of aggregation and detect subtle, non-obvious inconsistencies makes it a strong alternative to rigid rule-based checks. Beyond flagging anomalies, HrPCA offers diagnostic insights through eigenvector-based projections, enabling root-cause localization of upstream logging issues. In future work, we plan to evaluate this technique in real-world pipelines, explore GPU-accelerated versions, integrate active alerting systems, and build explainable dashboards that allow product and data engineers to trace anomalies to their origin quickly and confidently.


\begin{thebibliography}{9}

\bibitem{candes2011rpca}
E. J. Candès, X. Li, Y. Ma, and J. Wright,
\textit{Robust Principal Component Analysis?},
Journal of the ACM (JACM), vol. 58, no. 3, pp. 1–37, 2011.

\bibitem{zhou2010stable}
Zhou, Z., Li, X., Wright, J., Candes, E. J., and Ma, Y.,
\textit{Stable principal component pursuit},
IEEE International Symposium on Information Theory, 2010.

\bibitem{deequ}
Schelter, S., Biessmann, F., Rukat, T., and Schmidt, P.,
\textit{Automating large-scale data quality verification},
Proceedings of the VLDB Endowment, vol. 11, no. 12, 2018.

\bibitem{great_expectations}
Superconductive,\textit{Great Expectations: Always know what to expect from your data},
Available at: \url{https://greatexpectations.io/}

\bibitem{hrpca_background}
Zhou, T., and Tao, D.,
\textit{GoDec: Randomized low-rank and sparse matrix decomposition in noisy case},
ICML 2011.

\bibitem{anomaly_surveys}
Chandola, V., Banerjee, A., and Kumar, V.,
\textit{Anomaly detection: A survey},
ACM Computing Surveys (CSUR), 41(3), 2009.

\end{thebibliography}
\end{document}